\documentclass[twocolumn,
reprint
]{aastex62}

\renewcommand{\sc}{{\rm sc}}

\newcommand{\Wind}{\emph{Wind}}
\graphicspath{{./}{figures/}}

\shorttitle{Features of magnetic field Switchbacks observed near 1 au by \emph{\Wind}}

\begin{document}

%\title{Interpreting Solar Wind Data Beyond Taylor's Hypothesis}
\title{Features of magnetic field switchbacks in relation to the local-field geometry of large-amplitude Alfv\'enic oscillations: \emph{Wind} and \emph{PSP} observations}

%\title{\textcolor{red}{Magnetic switchback features observed by \emph{Wind} due to local field geometry: comparison with \emph{PSP} observations}}

\correspondingauthor{Sofiane Bourouaine}
\email{sofiane.bourouaine@jhuapl.edu}

\author{Sofiane Bourouaine}
\affil{Johns Hopkins Applied Physics Laboratory, 11100 Johns Hopkins Road, Laurel, MD, 20723, USA}

\author{Jean C. Perez}
\affil{Department of Aerospace, Physics and Space Sciences, Florida Institute of Technology, 150 W University Blvd, Melbourne, Fl, 32901, USA}

\author{Nour E. Raouafi}
\affil{Johns Hopkins Applied Physics Laboratory, 11100 Johns Hopkins Road, Laurel, MD, 20723, USA}

\author{Benjamin D. Chandran}
\affil{Department of Physics and Astronomy, University of New Hampshire, Durham, NH 03824, USA}

\author{Stuart D. Bale}
\affil{Space Sciences Laboratory, University of California, Berkeley, CA 94720, USA}

\author{Marco Velli}
\affil{Earth,Planetary, and Space Sciences, University of California, Los Angeles Los Angeles, CA 90095, USA}

\begin{abstract}
In this letter we report observations of magnetic switchback (SB) features near 1 au using data from the \emph{Wind} spacecraft. These features appear to be strikingly similar to the ones observed by the Parker Solar Probe mission (PSP) closer to the Sun: namely, one-sided spikes (or enhancements) in the solar-wind bulk speed $V$ that correlate/anti-correlate with the spikes seen in the radial-field component $B_R$. In the solar-wind streams that we analyzed, these specific SB features near 1 au are associated with large-amplitude Alfv\'enic oscillations that propagate outward from the sun along a local background (prevalent) magnetic field $\bf{B}_0$ that is nearly radial. We also show that, when $\bf{B}_0$ is nearly perpendicular to the radial direction, the large amplitude Alfv\'enic oscillations display variations in $V$ that are two-sided (i.e., $V$ alternately increases and decreases depending on the vector $\Delta\bf{B}=\bf{B} - \bf{B}_0$).
 As a consequence,  SBs may not appear always as one-sided spikes in $V$, especially at larger heliocentric distances where the local background field statistically departs from the radial direction. We suggest that SBs can be well described by large-amplitude Alfv\'enic fluctuations if the field rotation is computed with respect to a well determined local background field that, in some cases, may deviate from the large-scale Parker field.

%that relies mainly on highly Alfv\'enic oscillations that are associated with large rotation of the magnetic field (with rotational angle that may exceed 90$^\circ$) around a local prevalent field. in a such definition, the pattern of one sided enhancement of the bulk flow may not show up when the local prevalent field is perpendicular to the radial direction.

%Here we expect that, Based on the SBs features described above, SBs might appear with high statistics when the parker field is more radial (close to the sun), and gradually become less present when the parker field becomes less radial far from the sun. However, we define SBs as large-rotation magnetic field then we expect that the existence   

%This deviation might be due to the presence of large-scale (with period of days) large-amplitude waves  in the solar wind. According to this picture we would expect that SBs with that features appear with high statistics when the parker field is more radial (close to the sun), and gradually become less present when the parker field becomes less radial far from the sun.

\end{abstract}

\keywords{TBA}

\section{Introduction}

Near the Sun, at heliocentric distances below $\sim$0.17~au, \emph{Parker Solar Probe} \citep[\emph{PSP},][]{2016SSRv..204....7F} observes many intervals of slow wind where the bulk flow suddenly increases associated with temporary radial magnetic field reversals. These observed features were interpreted as magnetic switchbacks (SBs) %\citep[e.g., ][]{bale19,kasper19,dudok20} 
\citep[e.g., ][]{bale19,kasper19,2020ApJS..246...39D}.  Also, these observed SBs are characterized by a high degree of Alfv\'enicity. The word ``Alfv\'enicity'' is commonly related to the characteristics of Alfv\'en waves (with finite or small amplitude), which are an exact solution to 
the Magneto-hydrodynamics (MHD) equations \citep{walen44,goldstein74, barnes74}. These Alfv\'en waves are characterized by the following properties:  
%were found to satisfy the following criteria; 
1) they satisfy Wal\'en relations, $\Delta {\bf V}=\pm \Delta {\bf B}/\sqrt{\mu_0 \rho}$, where $\Delta {\bf V}$ and $\Delta {\bf B}$ are velocity and magnetic field perturbations around the background,
%the fluctuating velocity and magnetic fields, 
respectively, and $\mu_0$ is vacuum's magnetic permeability; 2) they have constant mass density, $\rho$, constant pressure, $p$, and constant magnetic field strength, $|{\bf B}|$; and 3) they propagate with a group velocity, called Alfv\'en velocity, ${\bf V_A}=\pm {\bf B}_0/\sqrt{\mu_0 \rho}$, i.e., either parallel (for $\Delta {\bf V}=- \Delta {\bf B}\sqrt{\mu_0 \rho}$ ) or anti-parallel (for $\Delta {\bf V}=+ \Delta {\bf B}\sqrt{\mu_0 \rho}$) to the local background magnetic field ${\bf B}_0$. Several in-situ measurements have revealed the existence of Alfv\'en waves in the solar wind that propagate mainly outward from the Sun \citep[see e.g., ][]{bavassano98}, 
%When Alfv\'en waves propagate outward from the Sun, 
which means the perturbed fields $\Delta {\bf V}$ and $\Delta {\bf B}$ either anti-correlate (when $B_{0R}>0$) or positively correlate (when $B_{0R}<0$). Here $B_{0R}$ is the component of ${\bf B}_0$ along the radial direction pointing outward from the Sun.

%(Bale et al. 2019; Kasper et al. 2019; de Wit et al. 2020; Horbury et al. 2020; Mozer et al. 2020; Rouillard et al. 2020; Tenerani et al. 2020). 
SBs have been previously observed over a wide range of heliocentric distances, \citep[see, e.g., ][]{borovsky16}, near 0.3 au  \citep{horbury18}, near 1 au \citep{kahler96,gosling09} and beyond 1 au \citep{balogh99,yamauchi04,neugebauer13}. 
In \emph{PSP} observations, the electron strahl pitch angle distributions was found to follow the magnetic field through SBs \citep{whittelsey20}.  
Also, within SBs, Alfv\'enic fluctuations at inertial-range scales appear to have correlations corresponding to fluctuations propagating towards the Sun \citep{bourouaine20,mcmanus20}.

Several scenarios have been proposed to explain the origin of SBs. Some studies suggest that SBs are caused by  magnetic reconnection as a result of interchange between open and closed magnetic field structures at the base of the solar corona, which are then convected outward with the solar wind \citep[e.g., ][]{fisk20,zank20,drake21,liang21}. Other scenarios propose that SBs are created locally in the solar wind as a result, for example, of the radial evolution of Alfv\'enic turbulence in the expanding solar wind \citep{landietal05,Landietal06, squire20,shoda21,mallet21} or due to shear-driven dynamics \citep[see e.g., ][]{Landietal06,Ruffolo20,schwadron21}.

 Unlike some previous works in which SBs are attributed to field deflections with respect to the large-scale Parker field \cite[see e.g.,][]{balogh99,borovsky16,matteini14}, in this letter we analyze the field rotation with respect to a local background field, that may deviate from the large-scale Parker spiral due to the presence of large-scale fluctuations (e.g., waves having periods of a few days \citet{coleman68}.)

In this letter we use \Wind~data to show that some of the SB features observed recently by \emph{PSP} near the Sun, such as the one-sided spikes (or enhancements) in the bulk flow that correlate/anti-correlate with the spikes seen in the radial component of the magnetic field,  are also observed near 1~au (where the large-scale Parker field is not nearly radial). Here, we demonstrate that such SBs features seem to show up naturally  when large-amplitude Alfv\'enic oscillations propagate anti-sunward along a nearly radial (or anti-radial) local prevalent field. Those SB features are also compared with the ones observed by \emph{PSP} during its first perihelion. In the following section, we present the analysis method and our findings. Then, in section 3, we summarize and discuss the obtained results.

%\vspace*{12pt}

\section{data analysis and results}

We use plasma and field measurements from \emph{Wind} to investigate the presence of SB features near 1 au. We use the combined data of magnetic field vector and the plasma parameters provided with time resolution of about 24.7~sec \citep{lepping95}. Here the vector fields are given in the Geocentric Solar Ecliptic system (GSE), i.e., the $x$--axis pointing from the Earth toward the Sun, the $y$--axis is chosen to be in the ecliptic plane pointing towards dusk (opposing planetary motion), and  the $z$--axis is parallel to the ecliptic north. For the sake of comparison with SBs observed by \emph{PSP}, we will also use plasma and magnetic field data from \emph{PSP} during its first encounter \citep{case20,bale16}. The \emph{PSP} data shown here are provided in the radial-tangential-normal (RTN) coordinate system. Here, we lower the resolution of \emph{PSP} data to the plasma resolution of \emph{Wind}, namely, 24.7 s for a better comparison.

\begin{figure}[!t]
    \centering
    \includegraphics[width=0.5\textwidth]{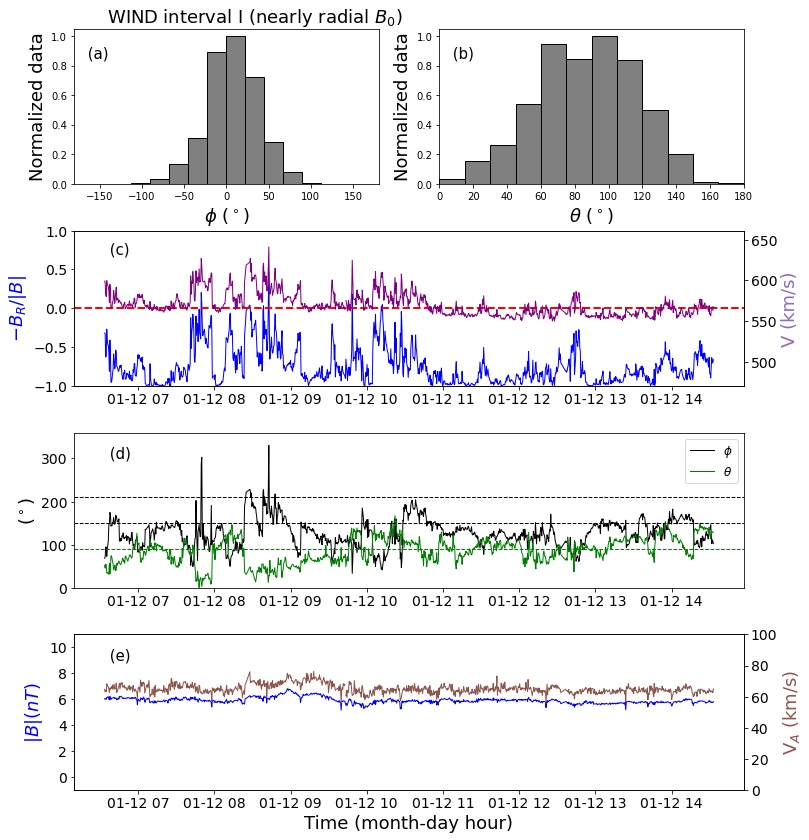}%{Figure1_switch.eps}
     \vspace*{-0.1cm}
    \caption{Eight-hours long time-interval I of \emph{Wind} observations: Panels from top to bottom correspond to (a) and (b) Histograms of the $\phi$ and $\theta$ angle, (c) the normalized radial component of the magnetic field $-B_R/|{\bf B}|$ (blue line) and the bulk flow $V$ (purple line), (d) The azimuth angle $\phi$ (black) and the polar angle $\theta$ (green), respectively, and (e) the magnitude of the magnetic field vector $|{\bf B}|$ (Blue) and the Alfv\'en speed $V_A$ (Brown).}
    \label{fig:fig1}
\end{figure}

\begin{figure}[!t]
    \centering
    \includegraphics[width=0.5\textwidth]{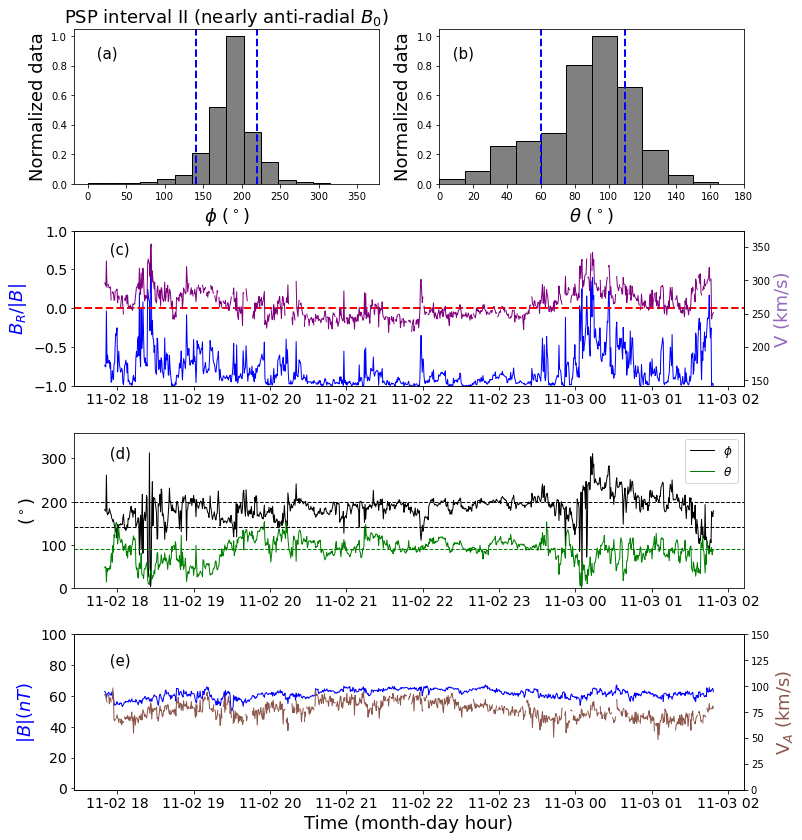}%{Figure1_switch.eps}
     \vspace*{-0.1cm}
    \caption{Eight-hours long time-interval II of \emph{PSP} observations: Panels from top to bottom correspond to (a) and (b) Histograms of the $\phi$ and $\theta$ angle, (c) the normalized radial component of the magnetic field $B_R/|{\bf B}|$ (blue line) and the bulk flow $V$ (purple line), (d) The azimuth angle $\phi$ (black) and the polar angle $\theta$ (green), respectively, and (e) the magnitude of the magnetic field vector $|{\bf B}|$ (Blue) and the Alfv\'en speed $V_A$ (Brown).}
    \label{fig:fig2}
\end{figure}

Figure~\ref{fig:fig1} displays results from an eight-hours long time-interval on January 12 in 2002 of \emph{Wind} measurements. Panels (a) and (b) show the histograms of the azimuth angle $\phi$ and the polar angle $\theta$, respectively. The azimuth angle $\phi$ is defined as the angle between the radial direction outward from the Sun ($-x$ direction) and the projected component of ${\bf B}$ onto the $x-y$ plane, and the polar angle $\theta$ is the angle between polar axis, $z$, and the instantaneous vector field ${\bf B}$. From the histogram plots, the most prevalent values of both angles correspond to $\phi\simeq 5^\circ$ and $\theta\simeq 86^\circ$. Here, the local mean field ${\bf B}_0$  (averaged over the eight-hours long time-interval) field is nearly radial (nearly parallel to the bulk velocity), and lies nearly on the ecliptic plane. 

%\textcolor{brown}{Ben:it might be useful to say something whether the observed wind streams are at or near the HCS}

Panel (c) of Figure~\ref{fig:fig1} displays the normalized x-component of the field $B_x/|{\bf B}|=-B_R/|{\bf B}|$ (where $B_R$ is the radial component) versus time. The normalized x-component $B_x=-B_R$ oscillates in a one-sided fashion, showing the kind of spikes that at some points can even exceed zero value and flip the sign. In the same figure we plot the flow speed $V$ (purple line) that also oscillates in one side fashion following the spikes in $-B_R/|{\bf B}|$. Panel (d) of Figure~\ref{fig:fig1} shows that the spikes in $B_x$ are mostly associated with the deviations of the angles $\phi$ and $\theta$ from their prevalent values.  

As shown in panel (e) of  Figure~\ref{fig:fig1} these field oscillations occur at nearly constant $B^2$ and constant Alfv\'en speed, $V_A=B/\sqrt{\mu_0 \rho}$, where $\rho$  is the mass density of protons. These are highly Alfv\'enic oscillations with a normalized cross helicity of $\sigma_c=-0.9$,   where $\sigma_c=2\langle (\delta {\bf V}\cdot \delta {\bf\hat{B}})/(\delta {\bf V}^2+\delta {\bf\hat{B}}^2)\rangle$, and $\delta {\bf V}={\bf V}- {\bf V_0}$ ($\delta {\bf \hat{B}}= ({\bf B}- {\bf B}_0)/\sqrt{\mu_0 \rho_0}$) is the fluctuating velocity field (magnetic field, converted to velocity unit). Here, ${\bf V_0}$ and $\rho_0$ are the mean quantities (averaged over the eight hours long time-interval) of the velocity and mass density, respectively. The Alfv\'enic structures measured in this interval propagate outward from the Sun (as $\sigma_c <0$ and $-B_{0x}=B_{0R}>0$).

%$\sigma_c=(E^{+}-E^{-})/(E^{+}+E^{-})$ with $E^{\pm}=|\delta {\bf v}\pm\delta {\bf b}|^2$  Here $\delta {\bf v}$ and $\delta {\bf b}$ are the fluctuating velocity and magnetic (in  velocity unit) fields, respectively.

%$E^{+}$ ($E^{-}$) is the energy of the %anti-sunward (sunward) propagating Alfv\'enic fluctuations.

The features shown in Figure~\ref{fig:fig1} appear to be similar to the ones observed by \emph{PSP} near the Sun at $\sim0.17$~au ~\citep[see, e.g.,][]{bale19,kasper19}. For the sake of comparison, in Figure~\ref{fig:fig2}, we plot the same parameters as those shown in Figure~\ref{fig:fig1} but from an eight-hour long time interval of \emph{PSP} during its first encounter on November 2 and 3 in 2018. Analysis of this interval shows that the local background field is nearly anti-radial corresponding to prevalent azimuth angle $\phi= 175^\circ$  (and $\theta= 88^\circ$). Here the $\phi$ angle is defined with respect to the radial axis of an RTN coordinate system, with $B_R \equiv -B_x$. The oscillations of the radial component $B_R$ correlate with the one sided enhancement in the bulk flow. These oscillations are also highly Alfv\'enic (with  $\sigma_c=0.9$) and propagate outward from the Sun  (as $\sigma_c>0$ and $B_{0R}<0$.)

\begin{figure}[!t]
    \centering
    \includegraphics[width=0.5\textwidth]{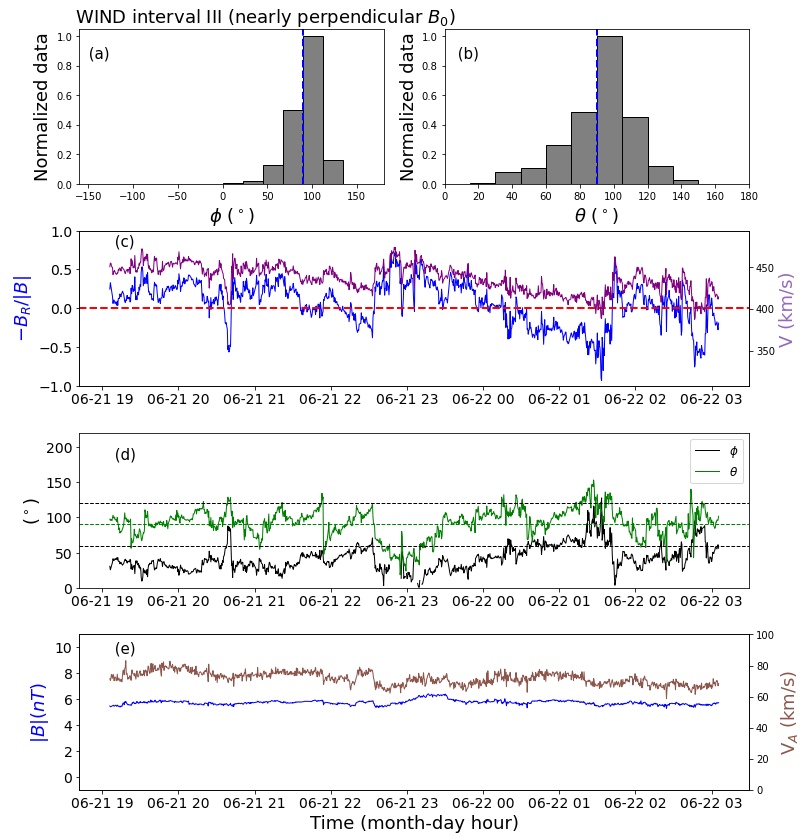}%{Figure1_switch.eps}
     \vspace*{-0.1cm}
    \caption{Interval III of \emph{Wind} observations: Panels from top to bottom correspond to (a) and (b) Histograms of the $\phi$ and $\theta$ angle, (c) the normalized radial component of the magnetic field $-B_R/|{\bf B}|$ (blue line) and the bulk flow $V$ (purple line), (d) The azimuth angle $\phi$ (black) and the polar angle $\theta$ (green), respectively, and (e) the magnitude of the magnetic field vector $|{\bf B}|$ (Blue) and the Alfv\'en speed $V_A$ (Brown).}
    \label{fig:fig3}
\end{figure}

In order to check how the geometry of the local background magnetic field may affect the overall features that are often associated with SBs near the Sun, we select another eight-hour long time interval observed by \emph{Wind} (shown in Figure~\ref{fig:fig3}) in which the oscillations are also highly Alfv\'enic (with $\sigma_c=-0.9$) and the local mean field is nearly perpendicular to the radial direction (and the solar wind bulk velocity). From the top panels of Figure~\ref{fig:fig3} we see that the most frequent value of $\phi$ ($\theta$) is $\sim 100^\circ$ ($\sim93^\circ$) corresponding to local background field ${\bf B}_0$ that is nearly perpendicular to the radial direction. In this case, the Alfv\'enic oscillations propagate anti-parallel to ${\bf B}_0$. 

For this geometrical configuration of the local field ${\bf B}_0$, the bulk flow and the radial component of the field do not show one-sided spikes as seen from Figures~\ref{fig:fig1} and~\ref{fig:fig2}.

\begin{figure}[!t]
    \centering
    \includegraphics[width=0.5\textwidth]{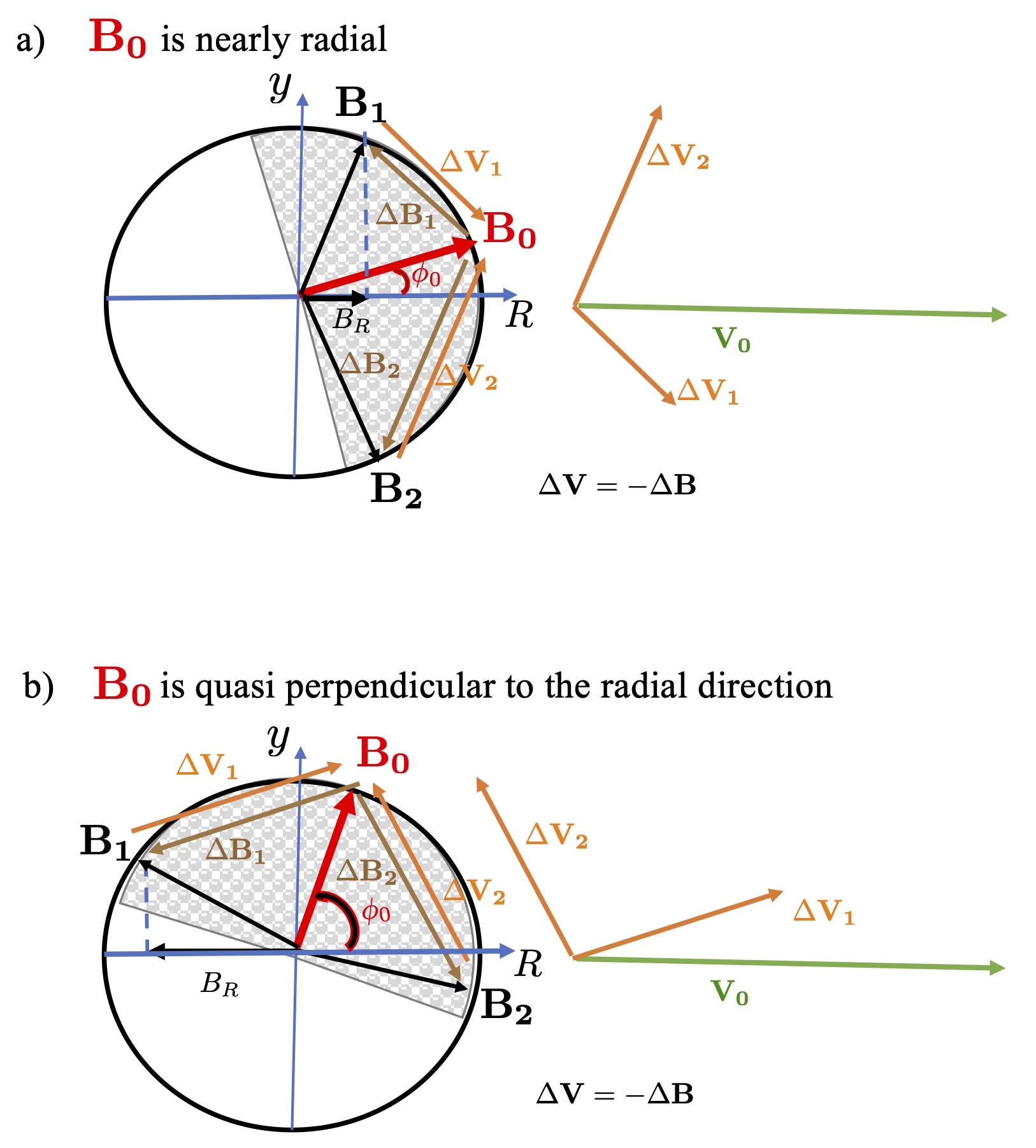}%{Figure1_switch.eps}
     \vspace*{-0.1cm}
    \caption{Sketch describes how the radial field component $B_x/|B|$ and the fluctuating bulk velocity ${\Delta{\bf  V_1}}= - \Delta {\bf B_1}$ and $\Delta {\bf V_2}= -\Delta {\bf B_2}$ (orange vectors) add to the local mean bulk velocity ${\bf V}_0$ (Green vector) during the oscillation of the instantaneous magnetic field vector ${\bf B}_1$ and ${\bf B}_2$ (black vectors) around the local background magnetic field vector ${\bf B}_0$ for the case (a) where ${\bf B}_0$ is nearly radial, and for case (b) where ${\bf B}_0$ is nearly perpendicular to the radial direction (of the bulk flow direction). Here $\Delta {\bf B_1}$ and $\Delta {\bf B_2}$ are normalized to velocity unit.}
    \label{fig:fig4}
\end{figure}

In the following we propose a picture, illustrated in Figure~\ref{fig:fig4}, to explain how the local background field direction and the corresponding anti-sunward Alfv\'enic oscillations ( in the case when ${\bf B}_0$ is nearly radial or anti-radial) may provide the above patterns of the bulk flow and the radial component of the field. For the sake of simplicity, we assume that the local background field ${\bf B}_0$  and the oscillating field vector ${\bf B}$ both lie nearly in the ecliptic plane. The demonstration below can still hold even when the oscillating field ${\bf B}$ is out of the ecliptic (i.e., with polar angle $\theta\ne 90$).

\begin{figure}[!t]
    \centering
    \includegraphics[width=0.5\textwidth]{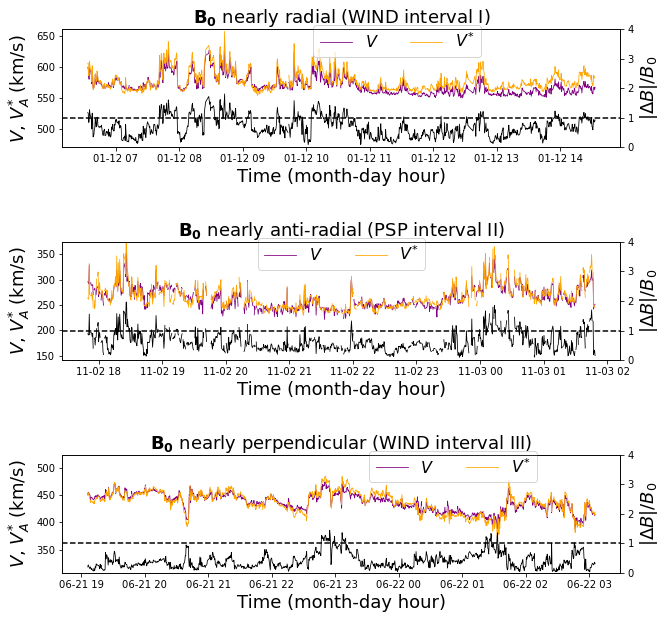}%{Figure1_switch.eps}
     \vspace*{-0.1cm}
    \caption{ Panels from top to bottom correspond to the measured bulk flow $V$ (purple line) and the reconstructed bulk flow $V^{*}$ from changes in magnetic field ${\Delta B}$ for interval I, II and III. The horizontal dash line represent the mean bulk flow ${\bf V}_0$ considered for the reconstruction of $V^{*}$.  }
    \label{fig:fig5}
\end{figure}

Figure~\ref{fig:fig4}a shows the case when the instantaneous magnetic field ${\bf B}$ (shown in black) oscillates in the shaded area around a nearly radial prevalent (background) magnetic field ${\bf B}_0$ (shown in red) (nearly parallel to the mean bulk flow velocity, ${\bf V}_0$, shown in green). If we assume that these oscillations are purely Alfv\'enic and propagate outward from the Sun, then the fluctuating magnetic field vector $\Delta {\bf \hat{B}}=({\bf B}-{\bf B}_0)/\sqrt{\mu_0 \rho}$  and the fluctuating bulk velocity vector $\Delta {\bf V}={\bf V}-{\bf V}_0$ have an opposite sign, i.e., $\Delta {\bf V}=-\Delta {\bf B}$ (in the unit where $\sqrt{\mu_0 \rho}=1$). Here ${\bf V}$ is the instantaneous measured bulk flow vector. In this case, because the magnetic field is nearly radial, the instantaneous field ${\bf B}$ oscillates around ${\bf B}_0$ within the shaded area, producing the vector changes $\Delta {\bf B}_1$ (counter-clockwise) and $\Delta {\bf B}_2$ (clockwise). In such rotations of $\bf B$ the radial component $B_R/|\bf B|$ (here $|\bf B|$ is constant) varies between positive and  negative values for some large rotations near or larger than $90^\circ$ leading to the kind of one-sided oscillations in $B_R$ that we see in the second panels of Figures~\ref{fig:fig1} and \ref{fig:fig2}. Interestingly, in this ${\bf B}_0$ configuration, it seems that the fluctuating velocity vectors $\Delta {\bf V}_1=-\Delta {\bf B_1}$ and $\Delta {\bf V}_2=-\Delta {\bf B_2}$ (orange vectors) both contribute to the enhancement of the bulk flow ${\bf V}_0$, leading to one-sided increases of the bulk flow that correlates with the spikes of $-B_R$. Note that this explanation holds even when ${\bf B}_0$ is nearly anti-radial and $\sigma_c>0$ for anti-sunward propagating Alfv\'enic oscillations, except that the one-sided increases of the bulk flow correlates with the spikes of $B_R$.

However, when the local background field ${\bf B}_0$ is nearly perpendicular to the radial direction (or to the bulk velocity) then the oscillating instantaneous field ${\bf B}$ around ${\bf B}_0$ can cover a wide range in which the radial component $B_R$  oscillates between positive and negative values in more or less equivalent ways. In such case, the $B_R$ component will not show a one sided pattern as shown in the second panel of Figure~\ref{fig:fig3}. Moreover, the counter-clockwise (clockwise) rotation of the field ${\bf B}$ with respect to ${\bf B}_0$ produces a change in the velocity $\Delta {\bf V}_1$ ($\Delta {\bf V}_2$) that leads to either increase (decrease) in the bulk velocity when $\sigma_c<0$ (or either decrease (increase) in the bulk velocity when $\sigma_c>0$) for such a ${\bf B}_0$ geometry.

To further verify the picture given above we estimate the fluctuating bulk velocity, $\Delta {\bf V^{*}}$, from the empirical fluctuating magnetic field $\Delta {\bf B}$ as $\Delta {\bf V^{*}}=\alpha \Delta {\bf B}$  (in the unit $\sqrt{\mu_0 \rho}=1$), where $\alpha=-1$ ($\alpha=1$) when ${\bf B}_0$ is near radial (anti-radial). This approximate estimation is based on the assumption that the oscillations are entirely Alfv\'enic and with anti-sunward propagation at least for the case when ${\bf B}_0$ is nearly radial or anti-radial.  Therefore, we estimate the modeled bulk flow, $V^{*}$, as $V^{*}=|{\bf V}_0+\Delta {\bf V^{*}}|$ for the three intervals we used above. Here  ${\bf V}_0$  is the mean bulk flow velocity.  Figure~\ref{fig:fig5} displays the velocity $V^{*}$, obtained from the model (orange line), and the bulk flow $V$ (purple line) obtained from direct measurements for the three intervals I (\emph{Wind}), II (\emph{PSP}) and III (\emph{Wind}) used in Figures \ref{fig:fig1}, \ref{fig:fig2} and \ref{fig:fig3}, respectively.
Figure~\ref{fig:fig5} shows that the modeled velocity, $V^{*}$ and the empirical bulk flow $V$ nearly overlap. It is clear that the enhanced bulk flow in most part of the signal in the two top panels can be well explained by the propagation of Alfv\'en waves along a nearly radial local background field ${\bf B}_0$ for the \emph{Wind} and the \emph{PSP} measurements. In addition, in the bottom panel of Figure~\ref{fig:fig5} we plot the ratio $\frac{|\Delta B|}{B_0}=2\sin(\eta/2)$, where $\eta$ is the rotational angle, i.e., the angle between the instantaneous field vector ${\bf B}$ and the local prevalent field ${\bf B}_0$. The figure clearly shows that the significant enhancement in the bulk flow occurs when $\Delta B/B_0 \gtrsim 1$ (one side spikes) when ${\bf B}_0$ is nearly radial (or anti-radial) (as shown for interval I and II). However, the bulk flow may significantly increase or decrease when $\Delta B/B_0 \gtrsim 1$ and ${\bf B}_0$ is nearly perpendicular.

\section{Summary and Discussion}
In this work we report observations of SB features near 1 au that are very similar to those observed by \emph{PSP} near the Sun. We have shown that a property often linked to the presence of SBs near the Sun, which is the one-sided spikes (or enhancements) in the bulk flow $V$ that correlate/anti-correlate with the spikes seen in the radial field component $B_R$, can be caused by the presence of large-amplitude  Alfv\'enic oscillations (with $|\Delta {\bf B}|/|\bf B_0|\sim 1$) propagating outward from the Sun along a local background field ${\bf B}_0$ that is nearly radial (or anti-radial). This property does not show up if the local prevalent field ${\bf B}_0$ is not sufficiently radial. 

\cite{matteini14} studied the dependence of solar wind speed on the magnetic field orientation in Alfv\'enic solar wind streams at high latitudes. The authors proposed that the enhancement in the bulk flow depends on the position of the instantaneous field {\bf B} with respect to a local mean field that is not assumed to be radial (but nearly follows Parker field direction at 1 au). 
In our analysis we rather focused on the geometry of the local background field, ${\bf B}_0$, of Alfv\'enic oscillations, and showed how that geometry affects the profile of the solar wind bulk flow when the field rotates strongly with respect to ${\bf B}_0$.

From our analysis, we conjecture that SBs (or at least a subset of SBs) can be sudden large rotations of the field (with $|\Delta {\bf B}|/B_0\sim 1$) associated with large-amplitude Alfv\'enic oscillations that propagate outward from the Sun along a well determined local background field, but this local field may in some cases deviate from the Parker spiral due to the presence of larger-scale solar wind oscillations, e.g., oscillations with periods of days \citep{coleman68,bruno05}. %Therefore, unlike in many previous works, here we suggest that SBs manifest as large-rotations of the field (with $|\Delta B|/B_0\sim 1$) with respect to the local prevalent field that is not necessary following the large-scale Parker field. 
Also, the one-sided spikes in the bulk flow speed that often appear to correlate/anti-correlate with the radial magnetic field component cannot be used as the main criterion for the determination of SB field reversals.

SBs (or at least a subset of them) are strongly connected to large-amplitude Alfv\'enic oscillations, which therefore makes  understanding the evolution of the large-amplitude Alfv\'enic fluctuations in the solar wind a necessary part of understanding the dynamics and evolution of SBs. For example, recent studies proposed that these type of oscillations can be generated due to the solar wind expansion \citep[e.g., ][]{squire20,mallet21}.

\acknowledgments
The authors would like to thank Dr. C. H. k. Chen for his valuable discussion.
SB was supported by NASA grants 80NSSCK0012 and Parker Solar Probe as part of of NASA’s Living with a Star (LWS) program (contract NNN06AA01C).
JCP was partially supported by NASA grant 80NSSCK0012 and NSF grant AGS-1752827. BC was supported in part by NASA grants NNN06AA01C and 80NSSC19K0829.
%Parker Solar Probe was designed, built, and is now operated by the Johns Hopkins Applied Physics Laboratory as part of NASA’s Living with a Star (LWS) program (contract NNN06AA01C). Support from the LWS management and technical team has played a critical role in the success of the Parker Solar Probe mission. SB's work was supported by grant \#

%\bibliography{MyLibrary.bib}

\end{document}